\documentclass{elsart3p}
\usepackage{epsf,graphicx,amssymb}
\journal{Powder Technology}
\begin{document}
\begin{frontmatter}

\title{Simulations of dense granular gases without gravity with impact-velocity-dependent restitution coefficient}
\author[sean]{Sean McNamara},
\author[eric]{Eric Falcon\thanksref{now}}
\thanks[now]{eric.falcon@univ-paris-diderot.fr}
\address[sean]{Institut f\"ur Computerphysik, Universit\"at Stuttgart, 70569 Stuttgart, Germany} 
\address[eric]{Mati\`ere et Syst\`emes Complexes, Universit\'e Paris Diderot--Paris 7, CNRS, 75 013 Paris, France}
 
\begin{abstract} 
We report two-dimensional simulations of strongly vibrated granular materials without gravity.  The coefficient of restitution depends on the impact velocity between particles by taking into account both the viscoelastic and plastic deformations
of particles, occurring at low and high velocities respectively.  Use of this model of restitution coefficient leads to new unexpected behaviors. When the number of particles $N$ is large, a loose cluster appears near the fixed wall, opposite the vibrating wall. The pressure exerted on the walls becomes independent of $N$, and linear in the vibration velocity $V$, quite as the granular temperature. The collision frequency at the vibrating wall becomes independent of both $N$ and $V$, whereas at the fixed wall, it is linear in both $N$ and $V$. These behaviors arise because the velocity-dependent restitution coefficient introduces a new time scale related to the collision velocity near the cross over from viscoelastic to plastic deformation. 
\end{abstract}
\end{frontmatter}

\section{Introduction}

A granular material is called a ``granular gas'' when 
the individual grains do not stay in contact with one another
but rather always move separately through space, interacting
only through dissipative collisions.  The absence of enduring contacts between the
particles allows granular gases to be treated by special numerical
and theoretical methods, such as event driven simulations and kinetic
theory.  These methods have given rise to a large body of knowledge
about dissipative granular gases \cite{GranularGases01,GranularGas03}.

The experimental realization of granular gases is however more difficult, because the grains must be continuously supplied with energy. This is usually done with a vibrating plate  \cite{Falcon:99}.  These experiments, however, often result in situations quite unlike those considered theoretically.  Some experiments have then been carried out in microgravity \cite{Falcon:99bis,Falcon:06} to limit the number of parameters in the problem in order to make easier the comparison with kinetic theory of dissipative granular gases. However, the interaction between experiments and theory has remained sporadic.

In our past work \cite{McNamara:05}, we have been able to compare simulations with experiments by studying a granular gas generated by placing grains in a box and then vibrating one of the walls to supply energy.  We have found that a velocity-dependent restitution coefficient is necessary to bring simulation and experiment into agreement.  In this paper, we investigate a further consequence
of this model that should be observable in vibrated granular gases in zero gravity.

To see why the velocity-dependent restitution coefficient has a radical effect in microgravity, one must enumerate the parameters describing the system.  The parameters are the following: the number of particles $N$, the diameter of the particles $d$, there mass $m$, the volume of the container, the restitution coefficient $r$, the vibration amplitude $A$, and vibration frequency $f$. Note that only one of these quantities -- the inverse of the vibration frequency, $1/f$ -- has the dimensions of time.  All other quantities are either dimensionless or involve length or mass.  Thus dimensional analysis can be used to determine the dependence of every quantity on $f$.  For example, both the average kinetic energy of the particles and the pressure vary as $f^2$.  However, this scaling is not in agreement with the one observed during microgravity experiments \cite{Falcon:99bis}.

There are two ways to disrupt this role of $f$.  First, one could
introduce gravity, bringing in a second time scale.  The second is to introduce a velocity dependent restitution coefficient.  As we show in this paper, this is sufficient to radically alter the behavior of the system.  Specifically, the coefficient is assumed to change its character at a specific value of the impact velocity $v_0$.  For velocities lower than $v_0$, collisions dissipate little energy, but above $v_0$, much energy is dissipated. This leads to several unusual features that could be observed experimentally.

\section{Parameters of the simulations}

\subsection{The variable coefficient of restitution}

The most important parameter in our simulations is the coefficient of
restitution.  The restitution coefficient $r$ is the ratio between the relative
normal velocities before and after impact.  In contrast to most previous
numerical studies of vibrated granular media
\cite{Luding:94a,Luding:94b,McNamara:98}, we let it depend on impact velocity.
In most simulations of strongly vibrated granular media, the
coefficient of restitution is considered to be constant and lower than 1.  

Dissipation during collisions of metallic particles can occur by two different mechanisms.  When the impact velocity $v$ is large ($v \gtrsim 5$ m/s \cite{Johnson:85}), the colliding particles deform fully plastically and $r \propto v^{-1/4}$, as reported experimentally \cite{Raman:1918,Tabor:48,Goldsmith:60,Labous:97} and theoretically \cite{Johnson:85,Tabor:48,Andrews:30,Thornton:97}. When $v\lesssim 0.1$
m/s \cite{Johnson:85}, the deformations are elastic with mainly viscoelastic
dissipation, and $1-r \propto v^{1/5}$, as reported experimentally \cite{Labous:97,Kuwabara:87,Falcon:98} and theoretically \cite{Kuwabara:87,Hertzsch:95}. Such velocity-dependent
restitution coefficient models have recently shown to be important in numerical
\cite{Brilliantov:00,Saluena:99,Bizon:98,Poschel:01,Goldman:98,Salo:88,Arsenovic:06} and experimental \cite{Falcon:98,Bridges:84} studies. Applications include:
granular gases \cite{McNamara:05}, granular fluid-like properties (convection \cite{Saluena:99}, surface waves \cite{Bizon:98}), collective collisional processes \cite{Falcon:98,Poschel:01,Goldman:98}, granular compaction \cite{Arsenovic:06}, and planetary rings \cite{Salo:88,Bridges:84}. 

\begin{figure}[tbp]
\centering
\includegraphics[width=.49\textwidth]{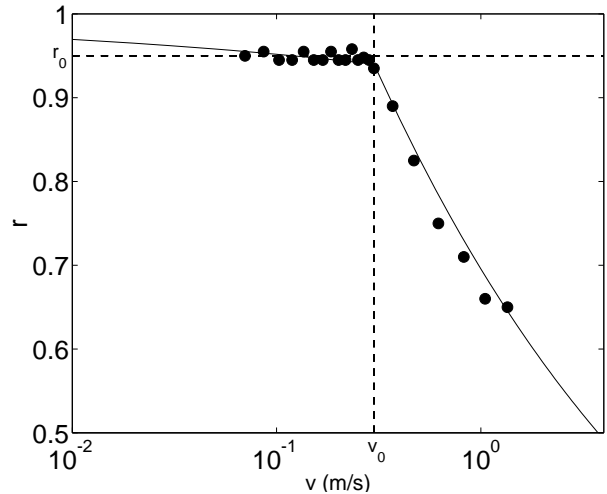}
\caption[]{The restitution coefficient $r$ as a function
of impact velocity $v$, as given in Eq.~(\ref{McNFal:eq:rvimp}) (solid line). 
 The dashed lines
 show $v_0=0.3$ m/s and $r_0=0.95$. Experimental points ($\bullet$) for
steel spheres were extracted from Fig.1 of Ref. \cite{Lifshitz:64}}
\label{McNFal:fig:rvimp}
\end{figure}
In this paper, we use a velocity dependent restitution coefficient  $r(v)$ and
join the two regimes of dissipation (viscoelastic and plastic) together as
simply as possible, assuming that
\begin{equation}
r(v) = \left\{  \begin{array}{ll} 
  1-(1-r_0) \left(\frac{v}{v_0} \right)^{1/5}, &  v \le v_0, \cr
     r_0
\left( \frac{v}{v_0} \right)^{-1/4}, & v \ge v_0 ,\end{array}\right.
\label{McNFal:eq:rvimp}
\end{equation}
where $v_0 = 0.3$ m/s is chosen, throughout the paper, to be the yielding
velocity for stainless steel particles \cite{Johnson:85,Lifshitz:64} for which
$r_0$ is close to 0.95 \cite{Lifshitz:64}. Note that $v_0 \sim 1/\sqrt{\rho}$
where $\rho$ is the density of the particle \cite{Johnson:85}. We display in
Fig.~\ref{McNFal:fig:rvimp} the velocity dependent restitution coefficient of
Eq.~(\ref{McNFal:eq:rvimp}), with $r_0=0.95$ and $v_0 = 0.3$ m/s, that agrees
well with experimental results on steel spheres from Ref.~\cite{Lifshitz:64}.
As also already noted by Ref.\ \cite{Johnson:85}, the impact velocity to cause
yield in metal surfaces is indeed relatively small.  For metal, it mainly comes
from the low yield stress value ($Y \sim 10^9$ N/m$^2$) with respect to the
elastic Young modulus ($E \sim 10^{11}$ N/m$^2$). Most impacts between
metallic bodies thus involve some plastic deformation. For more informations on restitution coefficient measurements, see Ref. \cite{Raman:1918,Tabor:48,Goldsmith:60,Labous:97,Kuwabara:87,Falcon:98,Bridges:84,Lifshitz:64,Foerster94}.

Note that other laws for the velocity dependent restitution coefficient have been studied in the context of rapid granular shear flows \cite{Turner:90}.  It was shown that such a coefficient changes the scaling relation between the imposed shear rate and the shear stress.  Specifically, when the restitution coefficient strongly decreases at high impact velocities, the pressure scales with the shear rate frequency with an exponent less than two.  This finding anticipates our results, but a detailed comparison is not possible because Ref.~\cite{Turner:90} does not use Eq.~(\ref{McNFal:eq:rvimp}).

\subsection{The other simulation parameters}
\label{sec:otherparams}

\begin{figure}[tbp]
\centering
\includegraphics[width=.49\textwidth]{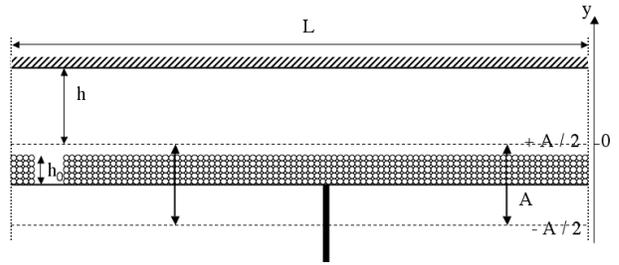}\\
\caption{Typical aspect ratio of the container for $n=5$ layers of particles (2 mm in diameter) leading to a height of the bed of particles at rest, $h_0=1$ cm. The container height is $h=2.5$ cm, its length $L=20$ cm, and the peak-peak vibrational amplitude $A=2.5$ cm (see text for details).}
\label{fig:AR}
\end{figure}

The numerical simulation consists of an ensemble of identical hard disks of
mass $m \approx 3 \times 10^{-5}$ kg excited vertically by a piston in a
two-dimensional box, in the absence of gravity (see Fig.~\ref{fig:AR}). We use the standard event-driven simulation method, where
collisions are assumed instantaneous and thus only binary collisions occur. To avoid inelastic collapse -- an unbounded number of
collisions in finite time \cite{McNamara:92} -- collisions
are made energy-conserving whenever very tight clusters of three particles
are detected.  Furthermore, note that using the restitution coefficient
given in Eq.~(\ref{McNFal:eq:rvimp}) prevents inelastic collapse \cite{Goldman:98}.
For
simplicity, we neglect the rotational degree of freedom.  Collisions with the
walls are treated in the same way as collisions between particles, except the
walls have infinite mass.  The simulation parameters are chosen close to
the usual ones used in the experiments (see for instance Ref.
\cite{Falcon:99}).  The particles are disks $d = 2$ mm in diameter with
stainless steel collision properties through $v_0$ and $r_0$ (see
Fig.~\ref{McNFal:fig:rvimp}). The vibrating piston at the bottom of the box has
amplitude $A=2.5$ cm, and frequencies 8 $\le f \le 30$ Hz.  The piston is
nearly sinusoidally vibrated with a waveform made by joining two parabolas
together \cite{McNamara:05}, leading to a maximum piston velocity given by
$V=4Af$ in the range $0.8 \leq V \leq 3$ m/s. For the parameters used in this paper, quantities such as the pressure are sensitive to $V$, but not to the maximum piston acceleration \cite{McNamara:05}, so $V$ will be used to characterize the vibration.

The box has width $L=20$ cm and horizontal periodic boundary conditions.  The number of layers of particles is $n=Nd/L$. Note that when $n<1$, it is also the fraction of the surface covered
by particles, so it could also be considered as an average surface fraction.
  A layer of particles, $n=1$, corresponds to $100$ particles.  We checked that $n$ is an appropriate way to measure the number of particles by also running simulations at $L=10$ cm and $L=40$ cm.  None of this paper's results depend significantly on $L$.  The height $h$ of the box depends on the number of particles in order to have a
constant difference $h-h_0=1.5$ cm, where $h_0$ is the height of the bed of particles at rest, $h$ being defined from the piston at its highest position (see Fig.~\ref{fig:AR}). $h-h_0$ is keep constant during most of the simulations (except when notified). All the simulations performed here are without gravity (except when notified).

\section{Results of simulations}

\subsection{Snapshots}

\begin{figure}
\centering
\includegraphics[width=.49\textwidth]{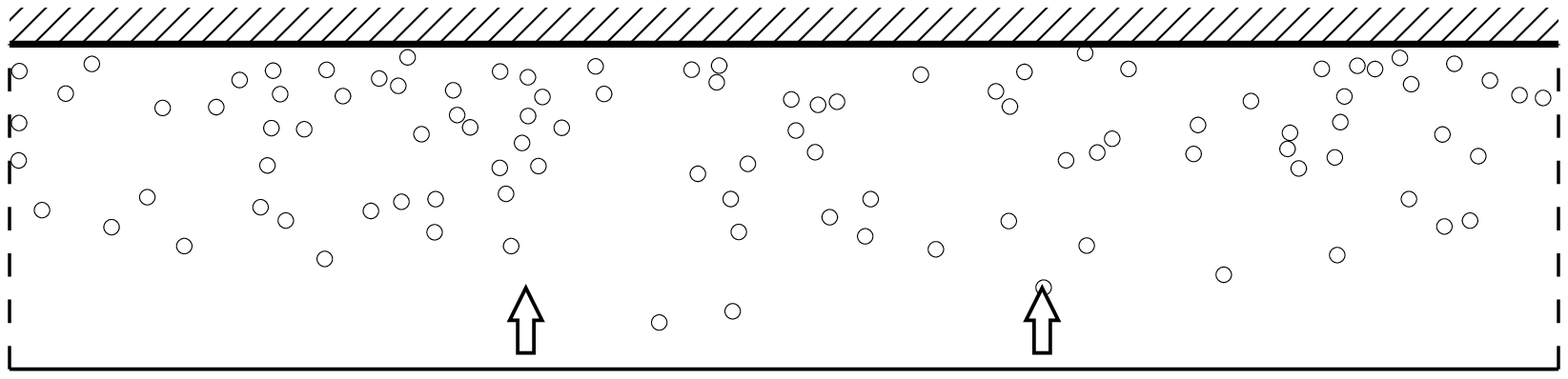}\\
\includegraphics[width=.49\textwidth]{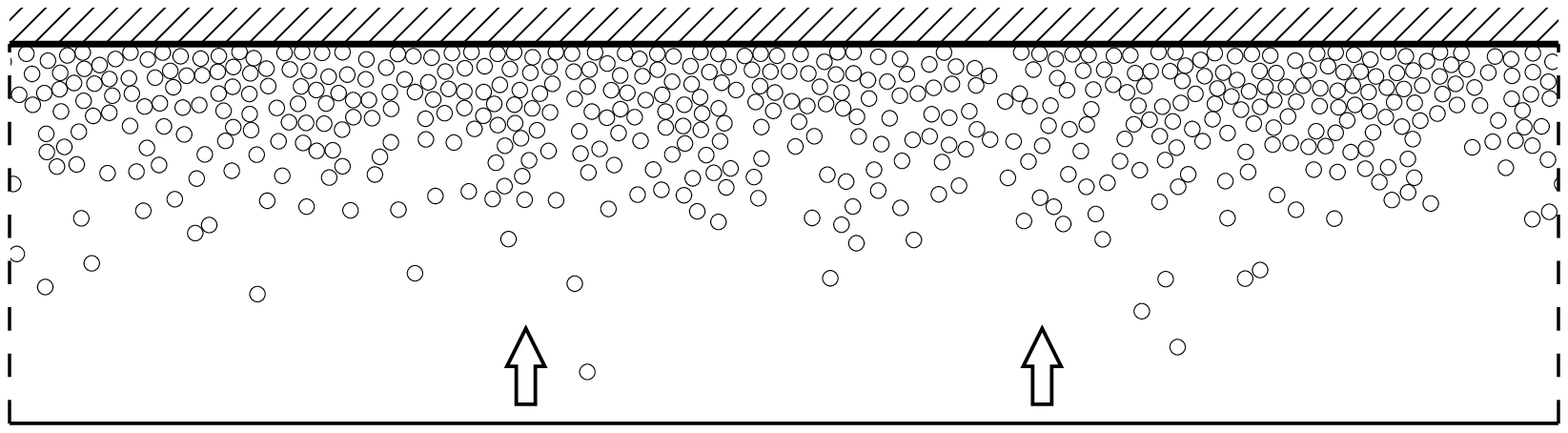}\\
\includegraphics[width=.49\textwidth]{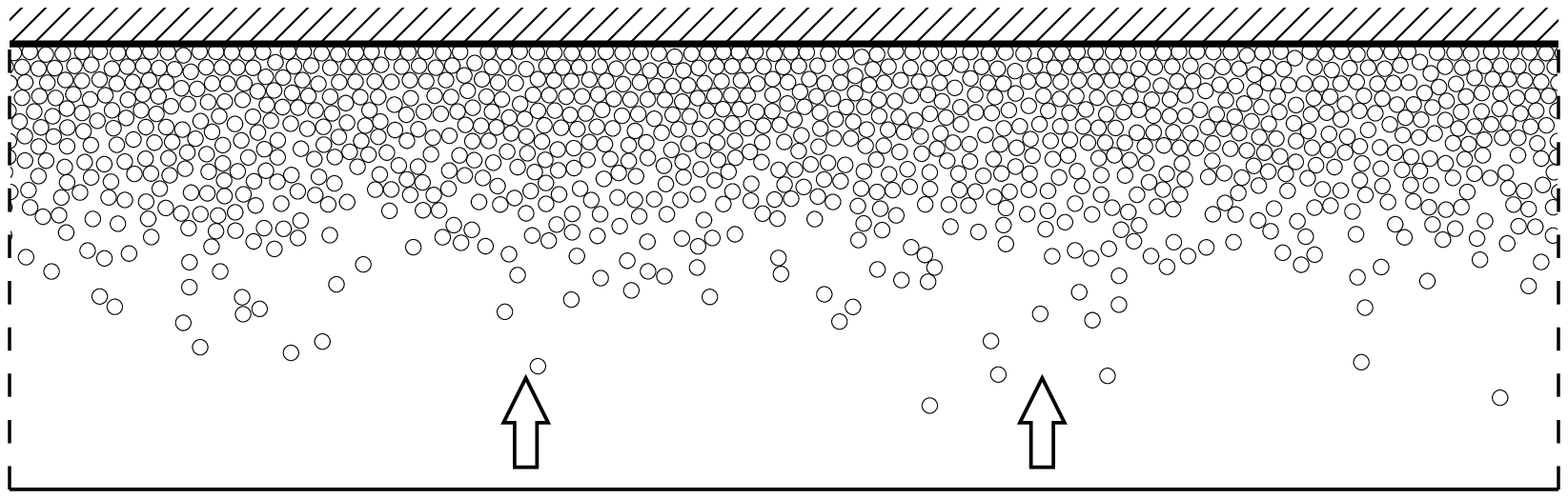}
\caption{Snapshots of various numbers of layers $n$: top $n=1$, middle: $n=5$,
bottom: $n=10$.  All snapshots taken when the vibrating wall (bottom) reaches
its lowest point.}
\label{fig:snapshots}
\end{figure}

The simplest way to display the results of the simulation is simply to show the
positions of the particles.  This is done in Fig.~\ref{fig:snapshots} for three simulations at
different particle numbers.  In all three panels, the wall shown at the top is
fixed, and the bottom, vibrating, wall is is shown at its lowest position.  The
periodic boundary conditions are shown by dotted lines at the sides of each
picture.  Note that through out the paper we will use words such as ``upper'',
``lower'', ``horizontal'', and ``vertical'' as suggested by the orientation of
these figures, although there is no gravity.

In the upper panel ($n=1$), the system merits the name of ``granular gas'': the
particles are well separated.  For the middle ($n=5$) and lower ($n=10$)
panels, the situation has changed.  A dense cluster forms against the
stationary wall.  As more particles are added, this cluster simply becomes
thicker.  Since the distance between the vibrating and stationary walls grows
with the number of particles, this process can continue indefinitely.

\subsection{The pressure}
\label{pressure}
\begin{figure}[tbp]
\centering
\includegraphics[width=.49\textwidth]{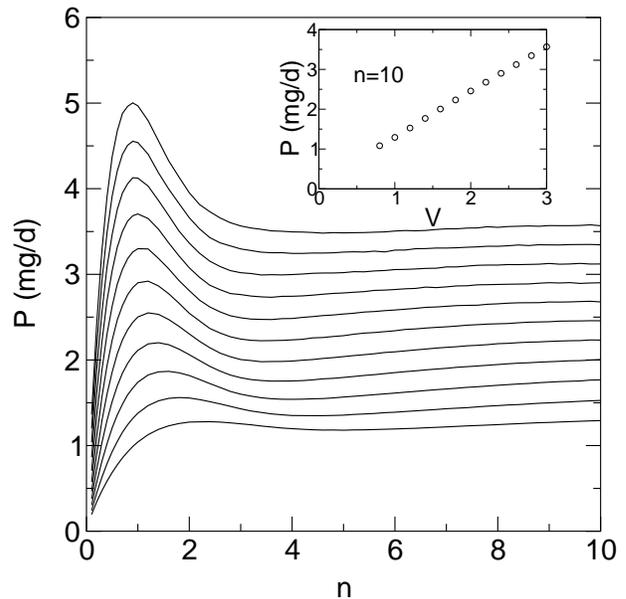}\\
\caption{Time averaged pressure $P$ on the top of the cell as a function of
particle layer, $n$, for various vibration velocities: $V=0.8$ to 3 m/s with
steps of 0.2 m/s (from lowest to uppermost curve). Inset: $P$ as a function of $V$ for $n=10$.}
\label{fig:Pn}
\end{figure}

Next we concentrate on the pressure on the upper wall (opposite the vibrating piston) since this is the quantity most accessible to experiments.  Here, the pressure is defined as the force per unit length that the particles exert on this wall, or alternatively the flux of
momentum, per unit time and length, through this wall.  Since the
collisions are instantaneously, a precise temporal resolution would yield a series of delta-function peaks.  We average the pressure over many ($6400$) cycles to obtain a stable average.  

We present in Fig.~\ref{fig:Pn} the dependence of $P$ on the piston velocity, $V$, and on the number of layers $n$.  This figure can be divided into two parts.  The dominant feature is the pressure peak observed near $n\approx1$.  A similar peak appears in the presence of gravity \cite{McNamara:05}. It is related to the increase of pressure with $n$ up to $n<1$ since interparticle collisions are rare and most of the particles are in vertical ballistic motion between the piston and the lid \cite{McNamara:05}. On
the other hand, for $n>3$, the pressure is approximately independent of the number of particles (see Fig.~\ref{fig:Pn}), and proportional to the piston velocity as shown by the inset of Fig.~\ref{fig:Pn}. The reason for this is discussed below.  Under gravity, the situation is quite different  \cite{McNamara:05}.  Adding particles causes more frequent interparticle collision, and the energy dissipation to increase and thus the pressure to decrease \cite{McNamara:05}.  At large values of $n$, resonances also appear under gravity.

\begin{figure}
\centering
\includegraphics[width=.49\textwidth]{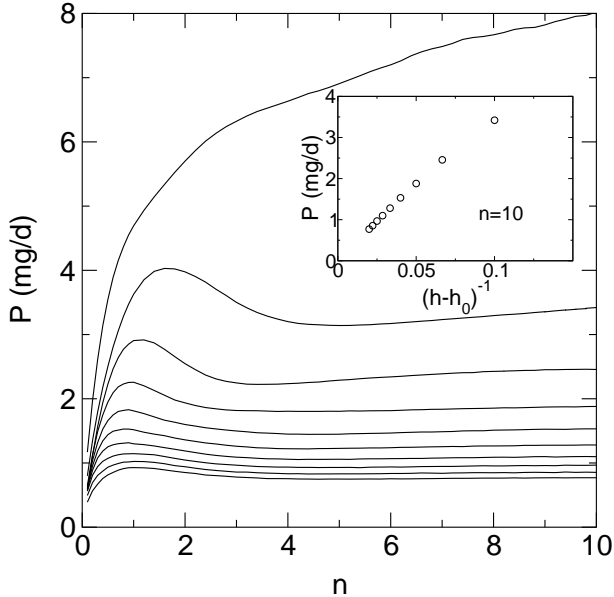}
\caption{Pressure as a function of $n$, for various heights of the cell: 0.5$\le h-h_0 \le
5$ cm with steps of 0.5 cm (from lowest to uppermost curve). Inset:
$P$ as a function of $(h-h_0)^{-1}$ for $n=10$.}
\label{fig:Phh0}
\end{figure}

In Fig.~\ref{fig:Phh0}, we examine how changing $h-h_0$ affects the pressure.
First, we note that the independence of the pressure on $n$ is not
confined to a special value of $h-h_0$, but holds for all values, except the
smallest ($h-h_0=0.5$ cm).  The pressure decreases as the height
increases.  Examining the dependence of the pressure on $h-h_0$ shows that $P
\propto (h-h_0)^{-1}$ as displayed in the inset in Fig.~\ref{fig:Phh0}.

\subsection{Scaling relations for global quantities}
\label{scalings}
We would now like to present the information presented in the previous sections
in a more compact way, and also consider other global quantities.  In addition to the
pressure, we will examine the granular temperature $T$, or mean kinetic energy per particle, and the collision frequency $C_\mathrm{up}$ of particles
with the upper wall.  From these quantities,
one can deduce the average impulse $\Delta I$ of a particle-wall collision,
$\Delta I_\mathrm{up} = P/C_\mathrm{up}$.

\begin{figure}[tbp]
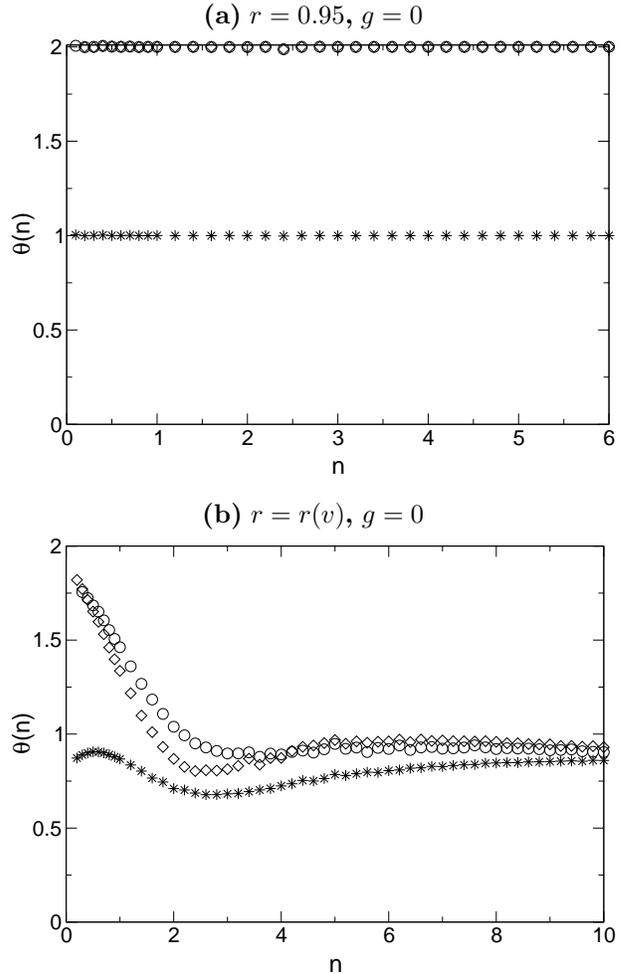

\centering
\begin{tabular}{c}
{\bf (a) $r=0.95$, $g=0$}\\
\includegraphics[width=.49\textwidth]{fig06a.eps}\\
{\bf (b) $r=r(v)$, $g=0$}\\
\includegraphics[width=.49\textwidth]{fig06b.eps}\\
\end{tabular}
\caption{Exponents  $\theta$ as a function of $n$ for simulations with {\bf (a)} $r=0.95$ or {\bf (b)} $r=r(v)$. The granular temperature, $T$ ($\Diamond$), collision frequency, $C_{\rm{up}}$ ($\ast$), and pressure, $P$ ($\circ$), scale like $V^{\theta(n)}$.}
\label{fig:exponents}
\end{figure}

The relation between the global quantities and the vibration velocity $V$
is reasonably well described by power-laws
\begin{equation}
\left.
\begin{array}{c}
T\\ C_\mathrm{up}\\ P
\end{array}
\right\}
 \propto V^{\theta(n)}
\label{eq:powerlaws}
\end{equation}
where the exponents $\theta(n)$ depend on the number of layers
\cite{McNamara:05}.  These exponents are shown in Fig.~\ref{fig:exponents}a for a constant restitution coefficient, and in Fig.~\ref{fig:exponents}b for a velocity-dependent restitution coefficient.

Each point in these figures was obtained by fixing $n$ and performing eleven
simulations, varying $V$ from 1 to 3 m/s.  The
amplitude $A$ is kept constant, and thus the vibration frequency is
proportional to $V$, as explained in Sec.~\ref{sec:otherparams}.  Then $\log X$
(where $X$ is the quantity being considered) is plotted against $\log V$.  The resulting curve is always nearly a straight line.  (Specifically $\left| \log X_\mathrm{observed}- \log X_\mathrm{fitted}\right|<0.1$ for all points). The slope of this line gives the power-law exponent $\theta$.

For the case of constant coefficient of restitution, $r=0.95$ (see Fig.~\ref{fig:exponents}a), the scaling exponents are independent of the number of particles: $C_\mathrm{up}\sim V^1$, $P$ and $T\sim V^2$.  As said above, this is precisely what is to be expected from dimensional analysis, since the vibration fixes the only time scale in the problem.  On the other hand, when $r=r(v)$, a more complicated behavior is observed (see Fig.~\ref{fig:exponents}b).  When $n$ is small, the exponents approach those of the previous case.  However, when the number of particles becomes large, all exponents approach unity.  The reason for this is discussed below.
 
\section{Anomalous scalings in the dense regime}
\label{McNFal:sec:g0clusters}
In Sects.~\ref{pressure} and \ref{scalings}, we have shown that, with a velocity dependent restitution coefficient, the pressure of a dense granular gas without gravity obeys the simple non-extensive relation
\begin{equation}
P \propto \frac{N^0 V^1}{h-h_0} \ {\rm .}
\label{McNFal:eq:Pscale}
\end{equation}
Let us now try to explain below this scaling. 

\subsection{Collision frequencies at the walls}
\begin{figure}[tbp]
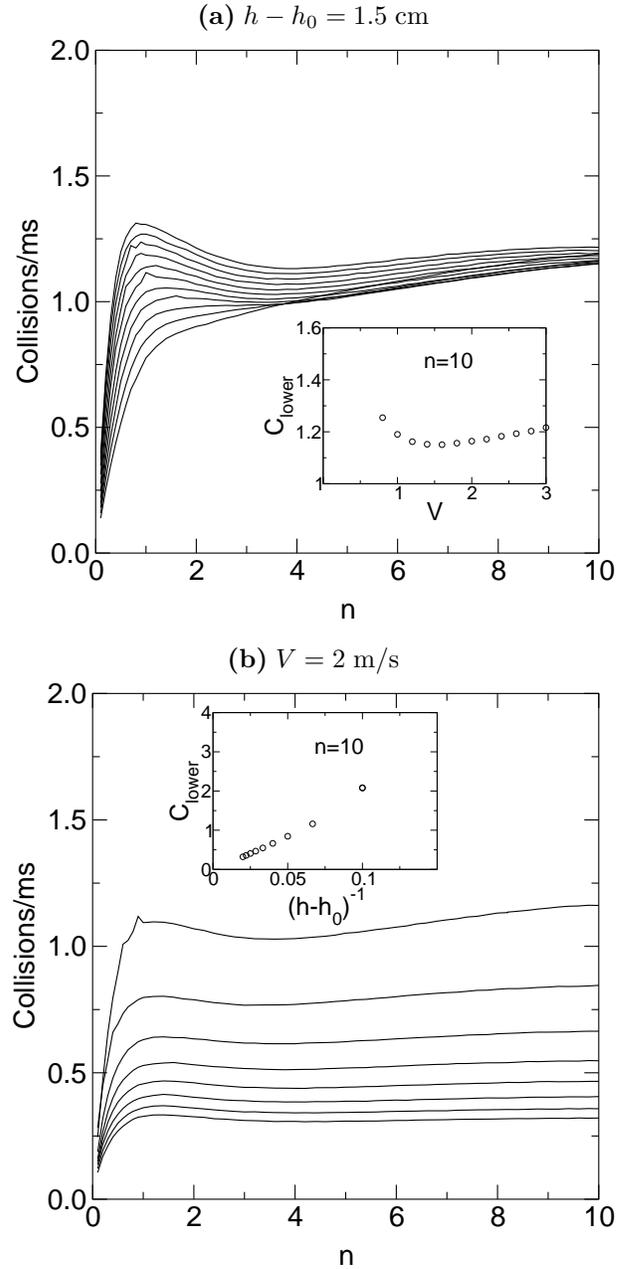

\centering
\begin{tabular}{c}
{\bf (a)} $h-h_0=1.5$ cm\\
\includegraphics[width=.49\textwidth]{fig07a.eps}\\
{\bf (b)} $V=2$ m/s\\
\includegraphics[width=.49\textwidth]{fig07b.eps}
\end{tabular}
\caption{The particle-piston collision rate $C_{\rm{low}}$ 
as a function of particle layer, $n$.
{\bf (a)} At fixed $h-h_0$, for various vibration velocities $V=1$ to 3 m/s with steps of 0.2 m/s  (from lowest to uppermost curve). Inset: $C_{\rm{low}}$ vs. $V$ for $n=10$.
{\bf (b)} At fixed $V$, for various heights $h-h_0=1$ to 5 cm with steps of $0.5$ cm (from lowest to uppermost curve). Inset: $C_{\rm{low}}$ vs. $(h-h_0)^{-1}$ for $n=10$.}
\label{McNFal:fig:Cdn}
\end{figure}

In Figs.~\ref{fig:snapshots}b-c, we observe that the majority of particles remain in a loose cluster pushed against the stationary wall, opposite the piston. Only those particles that ``evaporate'' from the cluster are struck by the piston.  The flux of evaporating particles can be estimated from the rate $C_{\rm{low}}$ of collisions between the piston and the particles.  This collision rate has a very curious behavior, as observed in Fig.~\ref{McNFal:fig:Cdn}.  $C_{\rm{low}}$ is roughly independent of $n$ when $n>3$ (see Fig.~\ref{McNFal:fig:Cdn}a). This behavior holds for other values of $h-h_0$ (see Fig.~\ref{McNFal:fig:Cdn}b). The inset of Fig.~\ref{McNFal:fig:Cdn}b shows that $C_{\rm{low}} \propto (h-h_0)^{-1}$. Moreover, the inset of Fig.~\ref{McNFal:fig:Cdn}a shows that $C_{\rm{low}}$ does not significantly  depend on $V$ (see the scale on the $y$-axes), and can be approximately considered as being independent of $V$. Thus, at high enough density, the collision frequency on the vibrating wall is found to be
\begin{equation}
C_{\rm{low}} \propto \frac{N^0 V^0}{h-h_0}  \ {\rm .}
\label{McNFal:eq:Cscale}
\end{equation}

\begin{figure}[tbp]
\centering
\includegraphics[width=.49\textwidth]{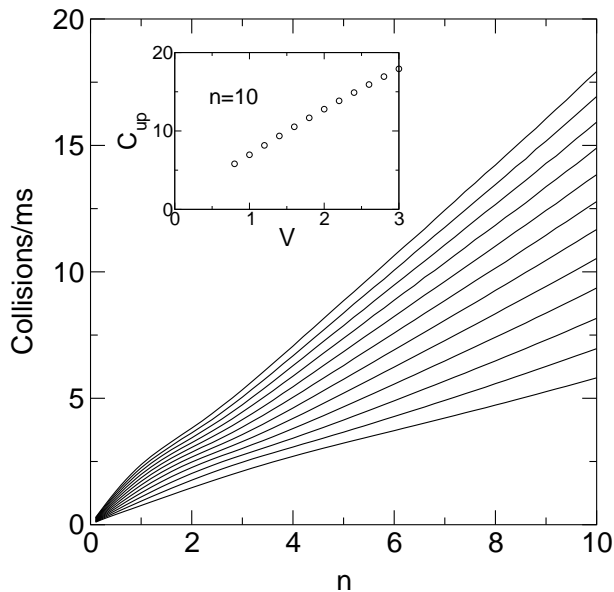}
\caption{$C_\mathrm{up}$ the collision frequency of particles with
the upper wall as a function of $n$ for for vibration velocities 
$V=0.8\,\mathrm{m/s}$ (lower curve) to $3\,\mathrm{m/s}$ (upper curve)
with steps of $0.2\,\mathrm{m/s}$.  Inset: $C_\mathrm{up}$ vs $V$ for
$n=10$.}
\label{fig:Cup}
\end{figure}

The rate $C_{\rm{up}}$ of collisions between the particles and the fixed wall is displayed in Fig.~\ref{fig:Cup}. It has a quite different behavior from $C_{\rm{low}}$. In the dilute limit ($n<2$), $C_{\rm{up}}$ increases more slowly than $n$ as already observed in microgravity experiments \cite{Falcon:06} and simulations \cite{Aumaitre:06}.  In the dense regime, when $n>3$, $C_{\rm{up}}$ is proportional to both the number of layers $n$ (see Fig.~\ref{fig:Cup}), and the piston velocity $V$ (see inset of Fig.~\ref{fig:Cup}). Thus, at high enough density, the collision frequency on the fixed wall is found to be 
\begin{equation}
C_{\rm{up}} \propto N^1V^1  \ {\rm .}
\label{McNFal:eq:Cup}
\end{equation}

\subsection{Explanation of the pressure scaling}
The time averaged pressure on a wall is the time averaged momentum flux divided by the area of a wall, that is
\begin{equation}
P_{\rm low} \propto C_{\rm low}\langle v_{\rm low}\rangle \ \ {\rm and}\ \  P_{\rm up} \propto C_{\rm up}\langle v_{\rm up}\rangle
\end{equation}
where $\langle v_{\rm low}\rangle$ and $\langle v_{\rm up}\rangle$ are, respectively, the mean particle velocities close to the piston and  close to the stationary wall. Since momentum is conserved, the flux of momentum entering the system at the piston, $C_{\rm low}\langle v_{\rm low}\rangle$, must have the same
value that the one leaving the system through the stationary wall, $C_{\rm up}\langle v_{\rm up}\rangle$. Therefore, the pressure on both sides is equal and is denoted $P$.

Figure~\ref{fig:snapshots}c shows that few particles are evaporated from the cluster, and are close to the piston. The evaporated particles from the cluster have the typical velocity of the particles within the cluster, that is $v_0$ (see below), and thus does not depend on the piston velocity. Moreover, there is no reason that the number of evaporated particles depends on $N$. Therefore, one can expected that $C_{\rm{low}} \propto  N^0V^0$ which is in agreement with our numerical results of Eq.~(\ref{McNFal:eq:Cscale}).

The scaling $C_{\rm{low}} \propto (h-h_0)^{-1}$ probably occurs
because a particle that evaporates from the cluster must
travel a certain distance before it encounters the piston.  This distance
increases with $h-h_0$ and thus the particle's travel time also increases. During its voyage, the evaporated particle could be struck by another particle that has just encountered the piston.  If this happens, both particles are scattered back into the cluster.  Thus the evaporated particle never reaches the piston.  If we assume that the probability of an evaporated particle being scattered back into the cluster is independent of time, the number of particles reaching the piston is inversely proportional to $h-h_0$.

When these evaporated grains collide with the piston, they acquire an upwards velocity proportional to $V$. Thus, the mean particle velocity close to the piston is $\langle v_{\rm low}\rangle \propto V$.   Therefore, using these two above results, one have $P_{\rm low} \propto C_{\rm low}\langle v_{\rm low}\rangle \propto N^0 V^1/(h-h_0)$ in agreement with our numerical results of Eq.~(\ref{McNFal:eq:Pscale}).

Similarly, the particles close to the fixed wall are within a cluster (see Fig.~\ref{fig:snapshots}c). Due to their numerous dissipative collisions within the cluster, these particles move little, even less that there are many particles within the cluster. Their mean velocity $\langle v_{\rm up}\rangle$ is thus fixed by the dissipation within the cluster (thus by $r(v)$ via $v_0$), and by the number of particles within the cluster (that is by $N$). Thus, close to the upper wall, one expect $\langle v_{\rm up}\rangle \propto v_0/N$. When the cluster is pushed against the upper wall, each layer contributes a fixed number of collisions.  Thus the number of collisions per cycle is proportional to $N$. The collision rate is also proportional to $V$, because the number of cycles per unit time grows linearly with $V$. Thus, one have $C_{\rm{up}} \propto  N^1V^1$ which is in agreement with our numerical results of Eq.~(\ref{McNFal:eq:Cup}). Therefore, using these two above results, one expect $P_{\rm up} \propto C_{\rm up}\langle v_{\rm up} \rangle \propto N^0 V^1$ in agreement with our numerical results of  Eq.~(\ref{McNFal:eq:Pscale}).

\section{Is the granular temperature relevant for dense granular gases?} 

In this section, we verified that the notion of granular temperature (mean kinetic energy per particle) is relevant in our dense system. Generally, for an homogeneous dilute granular gas, the pressure is proportional to the temperature. We wonder if the pressure scaling with the piston velocity, $P \propto V$, can be extended to the granular temperature for our dense system: $T \propto V$? Although our system is not spatially homogeneous and not stationary during a vibration cycle, we will see that the linear dependence of the granular temperature on $V$ is however meaningful as described below.

\subsection{Temporal distribution of energy}

\begin{figure}[tbp]
\centering
\includegraphics[width=.49\textwidth]{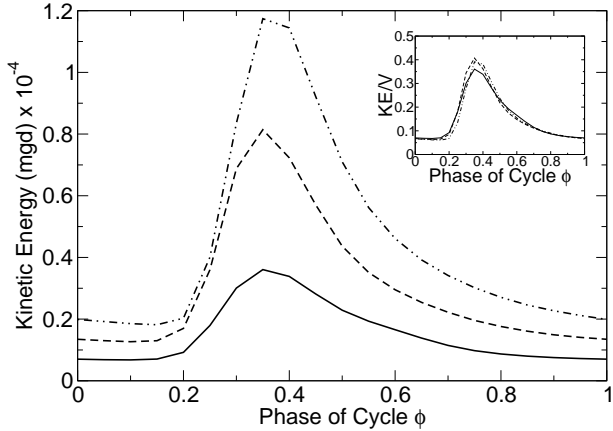}
\caption{Total kinetic energy of the particles as a function of vibration phase $\phi$ for three
different velocities $V=1$, 2 and 3 m/s (from lower to upper)
at constant number of particles $n=5$.  The wall is at
its lowest at $\phi=0$ and $\phi=1$, and reaches its highest point at
$\phi=0.5$.  Data from $100$ cycles were averaged to obtain these curves. Inset: Total kinetic energy rescaled by $V$ as a function of $\phi$.}
\label{fig:KE_time}
\end{figure}
We now examine the behavior of the simulations more closely, focusing on the variation of kinetic energy within one vibration cycle.  We define the ``phase'' of the vibration to be a number between $0$ and $1$ that gives the position of the vibrating wall.  When the wall is at its lowest position, $\phi=0$ or $\phi=1$.  When it has reached its highest position, $\phi=1/2$.  When it is halfway between its highest and lowest position, $\phi=1/4$ if it is ascending, $\phi=3/4$ if it is descending.

Fig.~\ref{fig:KE_time} shows the total kinetic energy of the particles as a function of the phase $\phi$, for three different piston velocities, all with the velocity dependent restitution coefficient.  Note that the kinetic energy varies varies by a factor of about six for each $V$.  The maximum occurs around $\phi\approx0.3$, just after the vibrating wall has attained its maximum velocity.
Considering the strong variations of kinetic energy during one cycle, one might
question whether the granular temperature $T$ of
the system were well-defined, or whether the law $T\propto P\propto V$ (see Eq.~\ref{McNFal:eq:Pscale}) is meaningful.

The law is meaningful, because the curves of Fig.~\ref{fig:KE_time} lie on one another if rescaled with $V$ as shown in the inset of Fig.~\ref{fig:KE_time}.  Thus, the granular temperature as a function of phase has the form $T(\phi) = g(\phi) V^1$, where $g$ is a function describing the shape of the curves in Fig.~\ref{fig:KE_time}.  If one measures $T$ at constant $\phi$ the same scaling law will be observed, independent of $\phi$. Note that to obtain the scaling exponents in Fig.~\ref{fig:exponents}, the granular temperature was measured $20$ times per cycle, and then averaged.

\subsection{Spatial distribution of energy}

\begin{figure}[tbp]
\centering
\includegraphics[width=.49\textwidth]{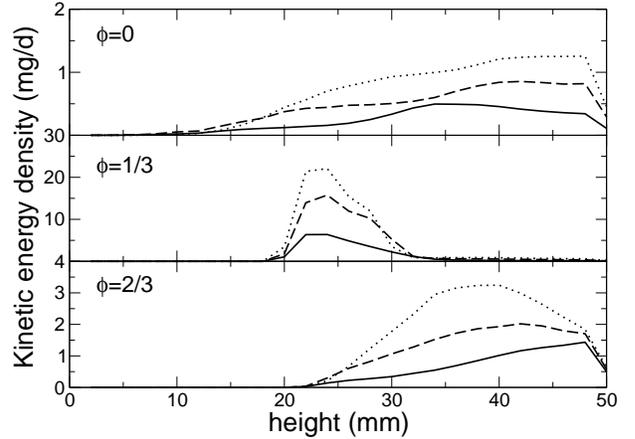}
\caption{Kinetic energy as a function of altitude $y$ at 3 different phases $\phi$ of the vibrating cycle. Each panel contains three curves corresponding to the three different $V$ in Fig.~\ref{fig:KE_time}. Data shown are for $V=1\,\mathrm{m/s}$ (lower curve), $V=2\,\mathrm{m/s}$ (middle curve),  and $V=3\,\mathrm{m/s}$ (upper curve), at constant number of layers $n=5$.}
\label{fig:KE}
\end{figure}

Under gravity, the altitude dependence of the density is usually measured (in order to extract the granular temperature). Far enough from the piston, the density decreases exponentially with altitude. A dense upper region supported on a fluidized low-density region near the vibrating piston is also reported experimentally \cite{Falcon:99}, numerically \cite{Lan:95} and predicted theoretically \cite{Kurtze:98}.

Without gravity, the spatial-dependence of the energy is examined in Fig.~\ref{fig:KE}.  These graphs were obtained by dividing the simulation domain into strips of height $2\,\textrm{mm}$, and then calculating the kinetic energy present in each strip.   Since the particles also have a diameter of $2\,\textrm{mm}$, each one will overlap two different strips.  A fraction of the particle's kinetic energy is assigned to these two strips, in proportion to the area of the particle located in each strip.  This procedure is carried out for all particles at fixed values of the phase $\phi$, and the results averaged over $100$ cycles.

The top panel shows the energy when the wall is at its lowest point
($\phi=0$).  At this phase, the energy is very low (note that the scales on the $y$-axes of
the three panels are all different), and its distribution resembles that of the
density.  In the second panel, the wall is now just past its maximum velocity.
There is a peak near $y\approx22$ mm, due to the kinetic energy just
injected by the wall.  This kinetic energy is ten times larger than the
kinetic energy found in the cluster, in spite of the small density. 
In the last panel, the wall has begun to move downward.  Note that in
all panels, the kinetic energy at any point is roughly proportional to
$V$.  Thus if one measures the kinetic energy density at fixed
$\phi$ and $y$ while varying $V$, one will observe a linear dependence
on $V$.

Figs.~\ref{fig:KE_time} and \ref{fig:KE} present a fairly complete
description of how energy flows through the system.  Energy is injected
for $0.2 \le \phi \le 0.4$, as the vibrating wall moves upward, colliding
with some particles that have escaped from the cluster found near the
upper wall.  These particles collide with this cluster at
$\phi\approx0.5$, exciting this cluster.  Then the energy decays,
so that by the time the wall begins moving upward again, most
of the energy has been dissipated.  

\subsection{Hydrodynamic and thermal kinetic energy}

\begin{figure}[tpp]
\centering
\includegraphics[width=.49\textwidth]{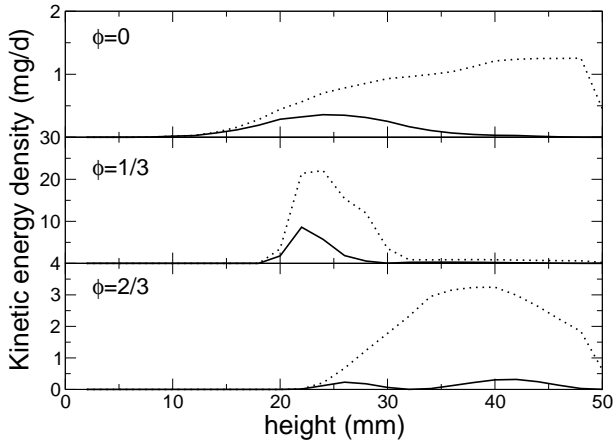}
\caption{Division of kinetic energy into mean motion ( ``hydrodynamic'') and disorganized parts ( ``thermal'') as a function of altitude at 3 different phases $\phi$ of the vibrating cycle. ($\cdots$) Thermal kinetic energy (data for $V=3$ from Fig.~\ref{fig:KE}).  ($-$) Hydrodynamic kinetic energy, obtained by
averaging the velocities of all particles found in each strip.}
\label{fig:KE_organized}
\end{figure}

In this section we consider the ratio of hydrodynamic to total
kinetic energy. Here, we use the word ``hydrodynamic'' as in the
context of granular kinetic theory.  It does not refer to any fluid
moving among the grains,
but concerns the decomposition of each grain's velocity into an
average ``hydrodynamic'' and a remaining ``thermal'' component.
The hydrodynamic velocity of a grain is found by averaging the
velocities of all nearby grains.  See Refs. \cite{Haff:83,Jenkins:85}
for a discussion on the distinction between these two energies in a granular medium.

The fraction of energy contained in the hydrodynamic velocities 
measures the organization of the flow.  If it is close to one, then
all the grains have nearly the same velocity.  If it is small,
the granular medium is in a gas-like state, where the randomized
motions of the particles dominate.
The use of the terms ``granular gas'' and ``granular temperature'' imply
that the granular medium is in a state like that of a usual gas: that
the ``thermal'' velocities of the grains are much larger than the
hydrodynamic ones.  But is this really the case? One could easily imagine a situation where a cluster of particles bounces back and forth between the two walls, without much relative motion between neighboring grains.

To see what situation applies to our simulations,
we return to the data used to produce Fig.~\ref{fig:KE}.  The
average velocity of the particles in each strip can be calculated, and
the kinetic energy related to this motion can be compared to the total.
The results are shown in Fig.~\ref{fig:KE_organized}.  The results show that near the piston, most of the energy is in the mean motion of the particles, even when the piston is descending ($\phi=2/3$). 
Thus the motion of the piston is supersonic.
Near the fixed wall, where most of the particles are located, however, most of the energy is thermal.  One can therefore conclude that in the cluster near the wall, gas-like conditions do prevail, i.e., most of the motion is thermal.  
Obviously, our dense granular gas differs from an usual gas in many other 
respects, such as the formation of cluster near the wall, and to the anomalous scalings reported here.

\section{Conclusions}
We report simulations of two-dimensional dense granular gas without gravity vibrated by a piston. The restitution coefficient used here depends on the relative velocity of particles. This allows to simulate a dissipative granular gas in a much more realistic way than using a constant restitution coefficient. This model of velocity dependent restitution coefficient is indeed in good agreement with experiments \cite{Raman:1918,Tabor:48,Goldsmith:60,Labous:97,Kuwabara:87,Falcon:98,Bridges:84,Lifshitz:64}. At high enough density, we observe a loose cluster
near the wall opposite to the vibrating one. This leads to unexpected scalings: the pressure, $P$, and the granular temperature, $T$, scale linearly with the piston velocity $V$. The collision frequency at the fixed wall and at the vibrating one scales respectively, as $N^1V^1$ and $N^0V^0$, where $N$ is the number of particles. We emphasize that these scalings can only be reproduced with this velocity dependent restitution coefficient.  If one uses
a constant restitution coefficient (as in most of previous simulations of granular gases), one obtains $P \propto T \propto V^2$ without gravity, no matter the constant value of the restitution coefficient. However, this $V^2$ scaling is not in agreement with the one reported during microgravity experiments in a dilute regime~\cite{Falcon:99bis}. Simulations of a dilute granular gas with velocity dependent restitution coefficient yield a scaling in agreement with this experimental one~\cite{McNamara:05}.

One difference between our simulations and the microgravity experiments on granular gases is that it is common to shake the whole container filled with particles in the experiments \cite{Falcon:99bis}.
One experiment has been recently performed by agitating dilute particles with a piston in low gravity \cite{Falcon:06}. The anomalous scalings, reported here numerically in a dense regime, may thus be observable in such microgravity experiments with many more particles. 
\bibliographystyle{elsart-num}
\bibliography{McNFal}

\end{document}